# Linear algorithms for phase retrieval in the Fresnel region: validity conditions

T.E. Gureyev[1,2,3] and Ya.I. Nesterets[2,3]

[1] *Monash University, Clayton, VIC 3800, Australia;*
[2] *Commonwealth Scientific and Industrial Research Organisation, Clayton, VIC 3168, Australia;*
[3] *University of New England, Armidale, NSW 2358, Australia.*

24 March 2015

## Abstract

We describe the relationship between different forms of linearized expressions for the spatial distribution of intensity of X-ray projection images obtained in the Fresnel region. We prove that under the natural validity conditions some of the previously published expressions can be simplified without a loss of accuracy. We also introduce modified validity conditions which are likely to be fulfilled in many relevant practical cases, and which lead to a further significant simplification of the expression for the image-plane intensity, permitting simple non-iterative linear algorithms for the phase retrieval.

## 1. Introduction

In recent years several results have been published [1-19] presenting various forms of linearized analytical expressions for the spatial distribution of the image-plane intensity in the case of in-line (projection) imaging (which involves free-space propagation of the transmitted wave from the exit surface of the object to the detector plane). The validity conditions under which the respective formulae can be derived have been discussed and analysed with varying degrees of rigor. No serious attempt seems to have been made so far to reconcile some of the "competing" expressions and compare theoretically their respective regions of validity. In the present paper we perform a detailed analysis of the validity conditions that were used explicitly or implicitly in previous publications and attempt to establish a definitive relationship between the respective results. We demonstrate that if the validity conditions required for their derivation are applied consistently, some of the formulae can be further simplified. The simplified expressions may also be more amenable to standard phase-retrieval approaches, where one collects one or more images in planes orthogonal to the optic axis at different object-to-detector distances, and then uses these images to retrieve the distribution of phase of the transmitted wave in the object plane. We then suggest a modified validity condition which is likely to be fulfilled in many relevant experimental arrangements, with the new condition leading to a particularly simple linearized expression for the image-plane intensity as a function of the object-plane phase. We also demonstrate that all of the considered linearized expressions reduce to the Transport of Intensity equation (TIE) in the limit of large Fresnel numbers, and they reduce to the first Born approximation (also known in this context as the weak object or Fourier Optics approximation) in the limit of weak absorption and small phase shifts. We hope that this exposition will help to clarify the relationship between the previously published results and will establish sufficiently clear validity conditions that could be used by researchers to determine the limits of applicability of various expressions under particular experimental conditions that may be encountered in the practice of phase-contrast imaging and tomography.

## 2. Guigay conditions and linearizability

Let an object (scatterer) be located in a vicinity of the optic axis in the half-space $z < 0$ immediately before the 'object' plane $z = 0$. We assume for simplicity that the wave incident on the sample is a plane monochromatic wave with wavelength $\lambda$ and unit intensity, propagating along the optic axis *z*, i.e. the complex amplitude of the incident wave is exp(*ikz*), $k = 2\pi/\lambda$. Generalization of the following results to cases involving polychromatic and spatially partially coherent incident radiation can be carried out similarly to the way described in reference [14]. The scattering properties of the object are assumed to be such that the wave transmitted through the object is paraxial, i.e. all the wavefront normals in the object plane are contained in a narrow cone around the direction of the *z* axis. The transmitted wave propagates in the free half-space *z* > 0 until it reaches a position-sensitive detector. As the transmitted wave has been assumed to be paraxial, its evolution in the free half-space *z* > 0 can be described by the Fresnel integral [1],

$$\mathbf{Fr}[q,R](x,y) = \frac{\exp(ikR)}{i\lambda R} \iint \exp\{\frac{i\pi}{\lambda R}[(x-x')^2 + (y-y')^2]\} q(x',y')\,\mathrm{d}x'\mathrm{d}y', \quad (1)$$

where $q(x,y) \equiv a(x,y)\exp[i\varphi(x,y)]$ is the complex scalar amplitude of the wave in the object plane and *R* is the distance between the object and image planes. The detector is assumed to be capable of measuring the spatial distribution of intensity in the image plane,

$$I_R(x,y) = |\mathbf{Fr}[q,R](x,y)|^2. \quad (2)$$

In phase-contrast imaging and phase-contrast tomography one is often interested in finding the object-plane phase $\varphi(x,y)$ and absorption[1] $\mu(x,y) = -\ln a(x,y)$ from the measured intensity distribution in one or more image planes $z = R_m$, $m = 1,2,..,M$. It is easy to see that eq. (2) is non-linear with respect to the object-plane phase and amplitude, and as such is usually rather challenging to solve analytically or numerically. Therefore, it appears useful to derive linearized forms (approximations) of eq. (2) which would be sufficiently accurate under certain well-specified conditions.

---

[1] It could be more appropriate to call this quantity "attenuation", rather than "absorption", as it usually also includes various scattering processes that lead to the reduction in the number of transmitted X-ray photons reaching the detector. We will use below the two terms interchangeably.

For simplicity, in what follows we mostly consider the one-dimensional situation (i.e. we omit the dependence of all functions on *y*). Generalizations of the derivations to the corresponding two-dimensional cases are straightforward and do not require any new insight. Some of relevant 2D formulae can be found in the next section of the paper.

The starting point for many known derivations of linear approximations to eq. (2) is the following expression for the Fourier transform of image intensity distribution given by Guigay in reference [2]:

$$\hat{I}_R(u) = \int \exp(i2\pi ux) q(x + \lambda Ru/2) \, q^*(x - \lambda Ru/2) \, dx, \qquad (3)$$

where $\hat{f}(u) = \int \exp(i2\pi ux) f(x) dx$ denotes Fourier transform and the superscript asterisk denotes complex conjugation. Equation (3) can be obtained directly by applying Fourier transform to the square modulus of (the one-dimensional version of) eq. (1). The following two assumptions were effectively employed in references [10,11,15] in order to linearize eq. (3) with respect to the object-plane phase distribution:

$$\varphi(x + \lambda Ru/2) - \varphi(x - \lambda Ru/2) = O(\varepsilon), \qquad (4)$$

$$a(x \pm \lambda Ru) - a(x) \mp \lambda Ru \, a'(x) = O(\varepsilon^2), \qquad (5)$$

where $\varepsilon << 1$ is a small (asymptotic) parameter, superscript prime sign denotes a derivative, and $O(\varepsilon)$ and $O(\varepsilon^2)$ denote quantities that are of the order of $\varepsilon$ and $\varepsilon^2$, respectively. Equation

(4) is known as Guigay's condition; it was first used in reference [2]. Equation (5) represents a form of linearizability condition for the real amplitude. Although it was not specified explicitly in references [10,11,15], one can verify that for the validity of the subsequent results it is sufficient to require that eqs. (4)-(5) hold for all $x$, such that $|x| < X_{max} + \lambda R u_{max}$, where $q(x) \equiv 1$ for $|x| \geq X_{max}$ [2], and for all $u$, such that $|u| \leq u_{max}$; here $u_{max} \equiv \min\{2U_{max}, u_{sys}\}$, where $U_{max}$ is the radius of the minimal circle enclosing the support of $\hat{q}(u)$ and $u_{sys}$ is the cut-off frequency of the imaging system (determined by its spatial resolution) (see e.g. [14, 16, 20])[3]. The reason for the existence of a particular upper limit on the required range of spatial frequencies in eqs. (4)-(5) can be easier appreciated from the following alternative form of eq. (3) which can be obtained by expressing $q(x)$ and $q^*(x)$ in eq. (3) via their Fourier transforms:

$$\hat{I}_R(u) = \int \exp(-i2\pi \lambda R u U) \hat{q}(U + u/2) \, \hat{q}^*(U - u/2) \, dU . \qquad (6)$$

It is obvious from eq. (6) that if $|u| > 2U_{max}$, then, for any $U$ either $\hat{q}(U + u/2) = 0$ or $\hat{q}^*(U - u/2) = 0$, and so $\hat{I}_R(u) = 0$.

It can be easily shown that when $u_{max} = \infty$, then eq. (4) implies that $\varphi(x) = C + \Delta\varphi(x)$, where $C$ is a constant and $\Delta\varphi(x) = O(\varepsilon)$ for all $x$ (if $q(x) \equiv 1$ for $|x| \geq X_{max}$, then $C = 0$).

Equation (4) is used to approximate eq. (3) with the help of the identity $\exp(\varepsilon) = 1 + \varepsilon + O(\varepsilon^2)$ applied to the phase:

---

[2] This setup corresponds to a finite object surrounded by completely transparent media; a complementary configuration, where the object is placed inside a finite aperture in an opaque screen, can be considered similarly.

[3] Strictly speaking, in the considered situation $q(x)$ cannot be band-limited, so formally $U_{max} = \infty$, and one should use the "essential support" [21] of $\hat{q}(u)$ in place of its true support. Note however that $u_{sys}$ is always finite, as the spatial resolution of an imaging system cannot be infinitely fine.

$$\hat{I}_R(u) = \int \exp(i2\pi ux) a(x + \lambda Ru/2)\, a(x - \lambda Ru/2)\, [1 + i\varphi(x + \lambda Ru/2) \\ - i\varphi(x - \lambda Ru/2)]\, dx + O(\varepsilon^2) = \hat{I}_R^{(0)}(u) + \hat{I}_R^{(+)}(u) + \hat{I}_R^{(-)}(u) + O(\varepsilon^2),$$ (7)

where the terms $\hat{I}_R^{(0)}$, $\hat{I}_R^{(+)}$ and $\hat{I}_R^{(-)}$ correspond respectively to the first, second and third terms in the square brackets under the integral. One can then apply equation (5) to $\hat{I}_R^{(\pm)}(u)$ and use the identity $i\lambda Ru \exp(i2\pi ux) = (R/k)[\exp(i2\pi ux)]'$ to obtain from eq. (7):

$$\hat{I}_R(u) \cong \hat{I}_R^{\varphi=0}(u) + 2\sin(\pi\lambda Ru^2)[I_0\varphi]^{\wedge}(u) + (R/k)\cos(\pi\lambda Ru^2)[(I_0'\varphi)']^{\wedge}(u).$$ (8)

The first term in eq. (8) corresponds to eq. (3) with $\varphi \equiv 0$. Equation (8) was originally obtained in reference [15].

With the use of eq. (5) it is possible to express
$\hat{I}_R^{\varphi=0}(u) = (i\lambda Ru/2)\sin(\pi\lambda Ru^2)[I_0']^{\wedge}(u) + \cos(\pi\lambda Ru^2)\hat{I}_0(u)$, which leads to the following equation, which was originally presented in reference [10]:

$$\hat{I}_R(u) \cong 2\sin(\pi\lambda Ru^2)\{[I_0\varphi]^{\wedge}(u) + (i\lambda Ru/4)[I_0']^{\wedge}(u)\} + \cos(\pi\lambda Ru^2)\{\hat{I}_0(u) \\ + (R/k)[(I_0'\varphi)']^{\wedge}(u)\}.$$ (9)

Thus, the linear eqs. (8) and (9) for the Fourier transform of the image intensity are equivalent to each other, in agreement with previous reports.

When the Fourier spectral power of the object-plane phase and intensity outside the low spatial frequencies, i.e. outside the region $\pi \lambda R U_{max}^2 << 1$, is of the order of $\varepsilon^2$, the sine function in eqs. (8) and (9) can be safely replaced by its argument and the cosine function can be replaced by 1 without a loss in accuracy order of the approximation. Under this condition one also obtains $\hat{I}_R^{\varphi=0}(u) \cong \hat{I}_0(u)$. Finally, the last term in eqs. (8) and (9) can be transformed into $(R/k)\{[(I_0\varphi)'']^{\wedge}(u) - [(I_0\varphi')']^{\wedge}(u)\}$. Therefore, under the condition $\pi \lambda R U_{max}^2 << 1$ eqs. (8) and (9) convert into the finite-difference form of the TIE [3]:

$$\hat{I}_R(u) \cong \hat{I}_0(u) - (R/k)[(I_0\varphi')']^{\wedge}(u). \tag{10}$$

In the case of phase shifts satisfying the Guigay condition, eq. (4), and small absorption, i.e. when $\mu(x) \equiv -\ln a(x) = O(\varepsilon)$, eqs. (8) and (9) transform into the first Born (Fourier Optics) type approximation [1, 2, 7]:

$$\hat{I}_R(u) \cong \delta(u) + 2\sin(\pi \lambda R u^2)\hat{\varphi}(u) - 2\cos(\pi \lambda R u^2)\hat{\mu}(u), \tag{11}$$

(here $\delta(u)$ is the Dirac delta-function) although this transformation is not easy to demonstrate. Instead, one could return to eq. (7) and use the identity $a(x + \lambda R u/2)\, a(x - \lambda R u/2) = 1 - \mu(x + \lambda R u/2) - \mu(x - \lambda R u/2) + O(\varepsilon^2)$, which holds when $\mu(x) = O(\varepsilon)$. This would allow one to derive eq. (11) from eq. (7) in the case of small absorption; eq. (5) is not required in this derivation. Note, however, that the smallness of the absorption does not imply eq. (5) in general. Therefore, the derivation of eq. (11) via eqs. (8)-(9) requires both eq. (5) and the condition $\mu(x) = O(\varepsilon)$ imposed together, even though eq.(11) follows from eq.(7) under the condition $\mu(x) = O(\varepsilon)$ alone.

Although it has not been spelled out in references [10,11,15] or elsewhere, one can see that it is indeed necessary to require that the right-hand side of eq. (5) is much smaller than $\varepsilon$ in order to discard the terms containing the second-order and higher derivatives of $a(x)$ in eqs. (8) and (9), because the terms of the order of $\varepsilon$ are retained explicitly in eq. (7) in accordance with eq. (4).

If the amplitude $a(x)$ satisfies Guigay's conditions similar to eq.(4), i.e. $a(x + \lambda Ru/2) - a(x - \lambda Ru/2) = O(\varepsilon)$ (which will be normally expected when the phase function satisfies eq.(4)), then it follows from eq.(5) that

$$\lambda Ru\, a'(x) = O(\varepsilon). \tag{12}$$

Equation (12) can be used to simplify the first term in eq. (8). It follows from eq. (5) directly that $a(x + \lambda Ru/2)\, a(x - \lambda Ru/2) = a^2(x) - (\lambda Ru/2)^2 (a')^2(x) + O(\varepsilon^2)$, and now we also have that $(\lambda Ru)^2 (a')^2(x) = O(\varepsilon^2)$, hence $a(x + \lambda Ru/2)\, a(x - \lambda Ru/2) = a^2(x) + O(\varepsilon^2)$, and the first term in eq. (8) can be replaced simply by $\hat{I}_0(u)$:

$$\hat{I}_R(u) \cong \hat{I}_0(u) + 2\sin(\pi\lambda Ru^2)[I_0\varphi]\hat{\,}(u) + (R/k)\cos(\pi\lambda Ru^2)[(I_0'\varphi)']\hat{\,}(u). \tag{13}$$

Thus we showed that eq. (13) is equivalent to eqs. (8) and (9) in most realistic cases, in the sense that all three of these equations provide an approximation of the same order of accuracy to the Fourier transform of the in-line (projection) image intensity, under the same conditions specified by eqs. (4) and (5), with the exception of truly "pathological" cases where the amplitude function does not satisfy the Guigay conditions despite eqs.(4) and (5), and the condition eq. (12) must be imposed independently and in addition to eq. (5), in order for eq.(13) to hold.

## 3. Symmetric complex Guigay conditions

Consider the Guigay condition for the absorption which is "symmetrical" to eq. (4) for the phase:

$$\mu(x + \lambda R u / 2) - \mu(x - \lambda R u / 2) = O(\varepsilon), \tag{14}$$

for all $x$ and all $u$, such that $|u| \leq u_{max}$.

It is possible to show that the Guigay condition for absorption, eq. (14), holds if and only if $\mu(x)$ can be represented as a sum

$$\mu(x) = \bar{\mu}(x) + \Delta\mu(x), \tag{15}$$

where $\bar{\mu}(x)$ is slowly varying, in the sense that $\lambda R u \, \bar{\mu}'(x) = O(\varepsilon)$, and $\Delta\mu(x)$ is small, i.e. $\Delta\mu(x) = O(\varepsilon)$, for all $x$ and all $u$, such that $|u| \leq u_{max}$ (of course, this applies equally to the Guigay condition for the phase, eq. (4)). Indeed, let us assume that eq. (14) holds and define $\bar{\mu}(x) = (2A)^{-1} \int_{x-A}^{x+A} \mu(t) dt$, where $A = 0.5 \lambda R u_{max}$. Then $\lambda R u \, \bar{\mu}'(x) = \lambda R u /(2A)[\mu(x + A) - \mu(x - A)] = O(\varepsilon)$ and $\Delta\mu(x) = \mu(x) - \bar{\mu}(x) = (2A)^{-1} \int_{x-A}^{x+A} [\mu(x) - \mu(t)] dt = O(\varepsilon)$ from eq. (14). Conversely, if eq. (15) holds, then $\mu(x + A) - \mu(x - A) = \bar{\mu}(x + A) - \bar{\mu}(x - A) + \Delta\mu(x + A) - \Delta\mu(x - A) = 2A \bar{\mu}'(\tilde{x}) + O(\varepsilon) = O(\varepsilon)$, which proves that eq. (14) holds as well.

Furthermore, if eq. (14) holds, it is always possible to find a representation in the form of eq. (15), where in addition to the above properties one also has $(\lambda Ru)^2 \bar{\mu}''(x) = O(\varepsilon)$ (consider e.g. $\bar{\mu}(x) = (2A)^{-2} \int_{x-A}^{x+A} \int_{t-A}^{t+A} \mu(s) ds dt$). However, it may not be always possible to choose the slowly varying component $\bar{\mu}(x)$ in such a way that $(\lambda Ru)^2 \bar{\mu}''(x) = O(\varepsilon^2)$, which would have implied that $\bar{\mu}(x)$ satisfies the linearizability condition, eq. (5) (see Appendix for a counter-example). Therefore, eqs. (4) and (14) do not imply in general that the phase and absorption functions can be represented as a sum of two components, one of which is of the order of $\varepsilon$ or smaller and the other is linearizable as above.

It is easy to verify that the "symmetric complex Guigay conditions" for the distributions of phase $\varphi(x)$ and absorption $\mu(x)$, as specified by eqs. (4) and (14) together, are equivalent to the following single condition for the complex amplitude $q(x) = \exp[-\mu(x) + i\varphi(x)]$:

$$q(x + \lambda Ru/2) - q(x - \lambda Ru/2) = q(x) O(\varepsilon), \qquad (16)$$

for all $x$ and all $u$, such that $|u| \leq u_{max}$. It is also possible to show that eq. (16) is equivalent to the condition that the function $q(x)$ can be represented as

$$q(x) = \bar{q}(x)[1 + \chi(x)], \qquad (17)$$

where $\bar{q}(x)$ is a slowly varying function in the sense that $\lambda Ru\, \bar{q}'(x) = \bar{q}(x) O(\varepsilon)$, and the function $\chi(x)$ is small, i.e. $\chi(x) = O(\varepsilon)$. To prove that, one can express any complex amplitude $q(x)$ as $q(x) = \exp[-\mu(x) + i\varphi(x)]$, define $\bar{q}(x) = \exp[-\bar{\mu}(x) + i\bar{\varphi}(x)]$, $\chi(x) = -\Delta\mu(x) + i\Delta\varphi(x)$, where $\bar{\mu}(x) = (2A)^{-1} \int_{x-A}^{x+A} \mu(t) dt$, $\bar{\varphi}(x) = (2A)^{-1} \int_{x-A}^{x+A} \varphi(t) dt$, $\Delta\mu(x) = \mu(x) - \bar{\mu}(x)$ and $\Delta\varphi(x) = \varphi(x) - \bar{\varphi}(x)$, and then use the same line of arguments as was used above for proving the equivalence of eq. (14) and eq. (15). It may not be possible in general to select the slowly varying component $\bar{q}(x)$ in the representation eq. (17) such that it would be linearizable in a sense similar to eq.(5).

The last point brings us to the result obtained in references [12, 14], where it was effectively assumed *a priori* that the object-plane complex amplitude can be represented in the form of eq. (17), but with $\bar{q}(x)$ being linearizable in the sense similar to eq. (5), i.e.

$$\bar{q}(x \pm \lambda Ru) - \bar{q}(x) \mp \lambda Ru\, \bar{q}'(x) = \bar{q}(x)O(\varepsilon^2), \tag{18}$$

for all $x$ and all $u$, such that $|u| \leq u_{max}$.

The conditions eq.(17)-(18) on the complex amplitude (including the requirement for the first derivative of $\bar{q}(x)$ to be small) was effectively used in references [12, 14] to obtain the following linearized form of eq.(3):

$$\hat{I}_R(u) \cong \hat{\bar{I}}_0(u) - (R/k)[(\bar{I}_0\bar{\varphi}')']^\wedge(u) - \\ 2\cos(\pi\lambda Ru^2)(\bar{I}_0\Delta\mu)^\wedge(u) + 2\sin(\pi\lambda Ru^2)(\bar{I}_0\Delta\varphi)^\wedge(u), \tag{19}$$

where $\bar{I}_0 = |\bar{q}|^2 = \exp(-2\bar{\mu})$. It is quite obvious that eq. (19) under the validity conditions expressed by eqs. (17)-(18) represents a combination of the TIE, eq. (10), for the slowly varying (linearizable) component, $\bar{q}(x)$, of the complex amplitude in the object plane, and a "correction" term given by the last two terms in eq. (19), which directly transform into the first Born type approximation, eq. (11), when $\bar{q}(x) \equiv 1$. For that reason, eq. (19) was called the "TIE+Born" approximation.

It follows from eq. (19) that for the validity of the TIE approximation, eq. (10), it is sufficient that $\chi(x) \equiv 0$ in eq. (17), i.e. $q(x) \equiv \overline{q}(x)$ is linearizable in the sense of eq. (18). Let us verify that the complex Guigay condition, eq. (16), together with the condition

$$\pi \lambda R u_{max} U_{max} \leq \varepsilon \ll 1, \qquad (20)$$

are in turn sufficient for the linearizability of the complex amplitude $q(x)$, and hence are also sufficient for the validity of the TIE. Using the Fourier convolution theorem, one can transform the obvious identity $q(x+A) - q(x-A) = d/dx \int_{x-A}^{x+A} q(t)dt$ into the following: $q(x+A) - q(x-A) = 2i \int \exp(i2\pi ux) \sin(2\pi uA) \hat{q}(u) du$, which is valid for an arbitrary $x$ and $A$. When $|A| \leq 0.5 \lambda R u_{max}$, then under the condition of eq. (20) one can approximate $\sin(2\pi uA) = 2\pi uA + O(\varepsilon^3)$ for any $u$, such that $|u| \ll U_{max}$. This allows the last identity to be re-written as $q(x+A) - q(x-A) = 2Aq'(x) + O(\varepsilon^3)$, which is the required linearization of the function $q(x)$ that can be used to derive the TIE from eq.(3). The smallness of the first derivative, $\lambda R u \, q'(x) = q(x) O(\varepsilon)$, which is also required for the derivation of the TIE, follows from the above linearization in conjunction with eq. (16).

It is possible to show by direct calculations that the TIE+Born approximation, eq. (19), is equivalent to eq. (13) (which in turn has been shown to be effectively equivalent to eqs. (8) and (9) in most realistic situations) if the eq. (5) holds in addition to the validity conditions specified by eqs. (17)-(18). Note that the "equivalence" is understood here in the sense that the difference between the quantities defined by eqs. (19) and (13) is of the order of $\varepsilon^2$ or smaller, as long as the validity conditions for these equations hold; in other words, eq. (13) and eq. (19) both represent correct approximations of the order of $\varepsilon^2$ to the general eq. (2). It is important to appreciate that eq. (5) does not follow from eqs. (17)-(18) in general, as eq. (5) demands the linearizability of the (whole) real amplitude $a(x) \equiv |q(x)|$, while eq. (18)

guarantees the linearizability of only the slowly varying component, $\bar{q}(x)$, of the complex amplitude $q(x)$. Conversely, eq. (18) does not follow from eq. (5), even in the presence of eq.(17), as eq.(18) effectively requires the linearizability of the slowly varying component of the phase function in addition to that of the real amplitude.

## 4. Asymmetric complex Guigay conditions

Thus, we have shown that the two linearized equations, eq. (13) and eq. (19), are not equivalent to each other in general, i.e. they generally require different conditions for their validity. It is clear, however, that the difference in the respective validity conditions is rather subtle, and in many practical situations may be negligible, i.e. both forms of the linearized equations may be valid at the same time. More importantly, the linearizability conditions, eqs. (5) and (18), required for the validity of these approximations, may be difficult to verify in an experimental situation where the properties of the imaged sample are usually not known *a priori* in sufficient detail. Therefore, it seems natural to look for a set of validity conditions that would have a good chance to hold under the experimental conditions typical for phase-contrast imaging and, at the same time, would allow one to use a linear approximation to the general expression for the image-plane intensity given by eq. (2). Recall, that the rationale for in-line phase-contrast imaging is usually the lack of sufficient absorption contrast, as is often the case for low-Z (e.g. biological) samples when imaged by high-energy X-rays. Note that the hard X-rays are used in order to insure a sufficient penetration power of the incident radiation, so that enough photons would be transmitted through the bulk of the specimen to form a projection image with acceptable noise statistics. Less absorption may also signify less radiation damage to the specimen as an added advantage of phase-contrast modes of imaging. The downside of using highly penetrating radiation for imaging low-Z samples is the weakness of the absorption contrast (as opposed to the total average absorption, which can be significant). When the so-called projection approximation is valid (see the discussion in Ref.[22,23]), which is typically the case in X-ray imaging of non-crystalline objects at spatial resolutions much coarser than the X-ray wavelength, then one can express the projected absorption and phase shifts as line integrals of the imaginary and real part, respectively, of the complex refractive index $n = 1 - \delta + i\beta$ of the sample, $\mu(x,y) = -2k \int_{-\infty}^{0} \beta(x,y,z) dz$ and $\varphi(x, y) = -k \int_{-\infty}^{0} \delta(x, y, z) dz$. As both $\beta$ and $\delta$ are usually non-negative everywhere and $\gamma = \beta / \delta \sim 10^{-2}$ - $10^{-4}$ in typical biomedical applications, the ratio $\mu(x, y) / \varphi(x, y)$ will be of the same order of magnitude. Therefore, when the Guigay condition eq. (4) holds for the phase, the corresponding absorption function is likely to satisfy a stronger condition:

$$\mu(x + \lambda Ru/2) - \mu(x - \lambda Ru/2) = O(\varepsilon^2). \tag{21}$$

We would like to emphasize here that under the conditions of eq. (21) the total absorption can be quite strong ($\mu(x) \sim 1$), and it is only the spatial variation of absorption that is assumed to be weak as a natural consequence of the Guigay condition imposed on the phase by eq. (4) and the assumption that we are dealing with hard-X-ray images of low-Z materials.

The linearized equations (8), (9) and (13) can be easily derived from eq. (3) under conditions specified by eq. (4) and eq. (21) (instead of eqs. (4)-(5)). In fact, these results can now be further simplified taking into account that
$a(x + \lambda Ru/2) = a(x) + O(\varepsilon^2) = a(x - \lambda Ru/2) + O(\varepsilon^2)$ under the conditions of eq. (21), leading to the following expression:

$$\hat{I}_R(u) \cong \hat{I}_0(u) + 2\sin(\pi \lambda Ru^2)[I_0 \varphi]^{\wedge}(u). \tag{22}$$

Let us show that eq. (19) can also be transformed into eq. (22) under the condition eq. (21). It is easy to verify that, when eq. (21) holds, the difference between the image-plane intensity $I_0(x)$ and its slowly varying component $\bar{I}_0(x) = \exp[-2\bar{\mu}(x)]$, where $\bar{\mu}(x) = (2A)^{-1} \int_{x-A}^{x+A} \mu(t) dt$, is of the order of $\varepsilon^2$, i.e. $I_0(x) = \exp[-2\mu(x)] = \exp[-2\bar{\mu}(x) + O(\varepsilon^2)] = \bar{I}_0(x)[1 + O(\varepsilon^2)]$. Hence, the term with $\Delta\mu$ in eq. (19) can be discarded in this case:

$$\hat{I}_R(u) \cong \hat{I}_0(u) - (R/k)[(I_0\overline{\varphi}')']^{\hat{}}(u) + 2\sin(\pi\lambda Ru^2)(I_0\Delta\varphi)^{\hat{}}(u). \qquad (23)$$

Furthermore, using the integration by parts one can obtain that
$-(R/k)[(\overline{I}_0\overline{\varphi}')']^{\hat{}}(u) = 2\pi\lambda Ru^2[\overline{I}_0\overline{\varphi}]^{\hat{}}(u) + O(\varepsilon^2) = 2\sin(\pi\lambda Ru^2)[\overline{I}_0\overline{\varphi}]^{\hat{}}(u) + O(\varepsilon^2)$ where we used eq. (21) and the properties of the slowly varying functions $\overline{I}_0(x)$ and $\overline{\varphi}(x)$. Finally, we can replace $\overline{I}_0(x)$ by $I_0(x)$ in the last expression and merge it with the last term in eq. (23), arriving at eq. (22). As a subtler point, note however, that in order to derive eq. (23) or eq. (19) under the "asymmetric complex Guigay conditions", eqs. (4) and (21), it is still necessary to assume the linearizability of the slowly varying component of the phase, $\overline{\varphi}(x)$.

When the Fourier spectra of the object-plane and image-plane intensities is non-zero only at low spatial frequencies, i.e. when $\pi\lambda RU_{max}^2 \ll 1$, the sine function in eq. (22) can be replaced by its argument leading to a reduced form of the TIE valid for objects with slowly varying absorption (as specified by eq. (21)): $I_R(x) \cong I_0(x)[1-(R/k)\varphi''(x)]$.

Equation (22) is linear with respect to the object-plane phase, and can be used for recovery of the phase if the intensity distributions in the object and image planes are known:

$$\varphi(x) = \frac{1}{I_0(x)} \mathbf{F}^{-1}\left[\frac{\hat{I}_R(u) - \hat{I}_0(u)}{2\sin(\pi\lambda Ru^2)}\right], \qquad (24)$$

where $\mathbf{F}^{-1}$ denotes the inverse Fourier transform. The singularities in eq. (24) (zeros of the sine function in the denominator) can be treated in a usual manner, e.g. by using the Tikhonov regularization [24] or more sophisticated techniques [25, 26].

Applications of the above linearized expressions for in-line image intensity to phase retrieval with real and simulated samples can be found in many publications, e.g.[2-20, 22-23, 25-30], including some containing comparative analysis of their accuracy. The present study is primarily concerned with the theoretical analysis of the validity conditions for previously published linearized approximations for in-line image intensity distributions, with the numerical simulations and experimental applications of these approximations having been reported elsewhere.

## 5. Summary

We have attempted to clarify the relationship between various forms of linearized equations in in-line (projection) imaging. We have shown that while eq. (9) obtained in references [10,11] and eq. (8) obtained in reference [15] are equivalent to each other and can be derived under the same assumptions about the object-plane phase and intensity, the alternative equation, eq. (19), obtained in references [12, 14] is not equivalent to the first two in the sense that its validity conditions are slightly different. On the other hand, under the superset of their validity conditions, all three forms of the linearized equations are equivalent, in the sense that they all represent a first-order (with respect to the small asymptotic parameter) approximation to the general non-linear expression for the image intensity given by the square modulus of the Fresnel integral. Finally, we have shown that in the cases typical for hard-X-ray imaging of low-Z samples, if the usual Guigay condition holds for the object-plane phase distribution, then the projected absorption distribution is likely to satisfy a stronger form of this condition, as given by eq. (21). Under such "asymmetric complex Guigay conditions" (eqs. (4) and (21)), one can obtain a very simple linear expression (eq. (22)) for the object-plane intensity, which can be conveniently employed for non-iterative phase retrieval.

**Appendix**

Here we present a counter-example disproving the following conjecture.

Conjecture: any function $\mu(x)$ satisfying the Guigay condition,

$$\mu(x+h) - \mu(x-h) = O(\varepsilon), \qquad (A1)$$

for all $x$ and all $h$, such that $|h| \leq A$, can be represented as a sum $\mu(x) = \bar{\mu}(x) + \Delta\mu(x)$, where $\Delta\mu(x)$ is small, i.e. $\Delta\mu(x) = O(\varepsilon)$, and $\bar{\mu}(x)$ is linearizable, i.e. $\bar{\mu}(x \pm 2h) = \bar{\mu}(x) \pm 2h\,\bar{\mu}'(x) + O(\varepsilon^2)$ for all $x$ and $h$, such that $|h| \leq A$, and $2h\bar{\mu}'(x) = O(\varepsilon)$.

Intuitively, it is appealing to represent the slowly varying and the small components in a parametric form as $\bar{\mu}(x) = \mu_1(0.5\varepsilon\,x/A)$ and $\Delta\mu(x) = \varepsilon\mu_2(x)$, where $\mu_1$ and $\mu_2$ are some bounded functions. In this case it is easy to see that $2h\,\bar{\mu}'(x) = \varepsilon(h/A)\mu_1'(0.5\varepsilon\,x/A) = O(\varepsilon)$ for $|h| \leq A$ and $\Delta\mu(x) = O(\varepsilon)$ as required. Moreover, $(2h)^2\,\bar{\mu}''(x) = \varepsilon^2(h/A)^2\mu_1''(0.5\varepsilon\,x/A) = O(\varepsilon^2)$, and hence $\bar{\mu}(x \pm 2h) - \bar{\mu}(x) \mp 2h\,\bar{\mu}'(x) = 2h^2\,\bar{\mu}''(\tilde{x}) = O(\varepsilon^2)$, which then also implies that $\bar{a}(x) = \exp[-\bar{\mu}(x)]$ is linearizable in the sense of eq. (5). However, the representation $\mu(x) = \mu_1(0.5\varepsilon\,x/A) + \varepsilon\mu_2(x)$, where $\mu_1$ and $\mu_2$ are some bounded functions, does not have to exist for every function satisfying eq. (A1).

For an arbitrary function $\mu(x)$ satisfying the Guigay condition (A1) let us define $\Delta\mu(x) = \mu(x) - \bar{\mu}(x)$, where $\bar{\mu}(x)$ is a piecewise-linear approximation to $\mu(x)$, i.e. $\bar{\mu}(x) = \sum_{n=-\infty}^{\infty}[\mu(An) + (x - An)f_n]\chi_{[An, An+A]}(x)$, where $f_n = [\mu(An+A) - \mu(An)]/A$, $\chi_{[a,b]}(x)$ is equal to 1 when $a \leq x \leq b$, and $\chi_{[a,b]}(x)$ is equal to 0 otherwise. Then it is easy to see that $\Delta\mu(x) = O(\varepsilon)$ as a direct consequence of eq. (A1). Therefore, it is sufficient to prove the Conjecture just for piecewise-linear functions $\bar{\mu}(x)$ whose values at the apex points $An$ satisfy eq. (A1). For that, it would be sufficient to approximate any such piecewise-linear functions within the strip $[\tilde{\mu}(x) - \varepsilon, \tilde{\mu}(x) + \varepsilon]$ by a smooth function with the second derivative of the order of $\varepsilon^2/A^2$. Let us show that this is <u>not</u> always possible.

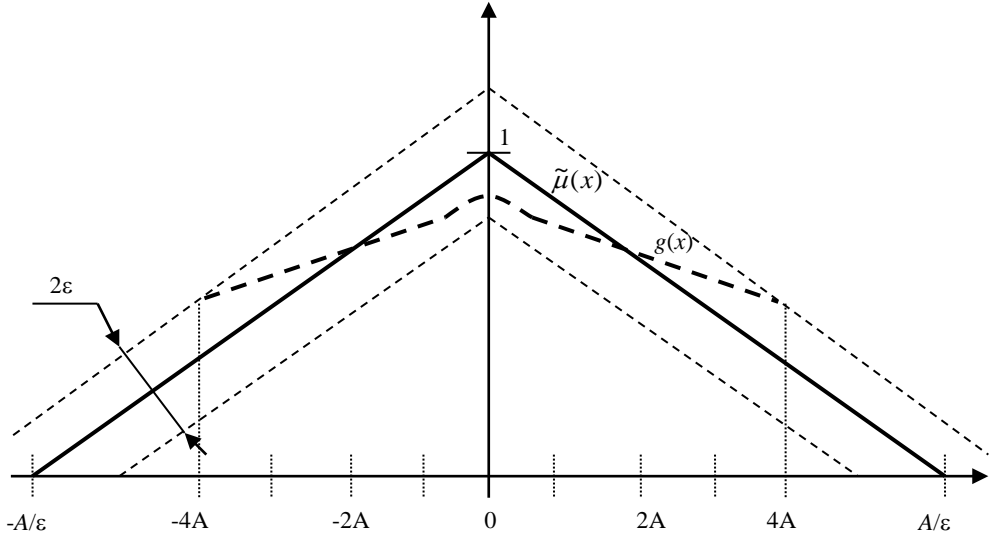

Fig. A1. Counter-example for the Conjecture.

Let $\tilde{\mu}(x) = (\varepsilon/A)(x + A/\varepsilon)\chi_{[-A/\varepsilon,0)}(x) + (\varepsilon/A)(-x + A/\varepsilon)\chi_{[0,A/\varepsilon]}(x)$ (Fig.A1), which is an example of a piecewise-linear function $\bar{\mu}(x)$ of the type described above. If a smooth function $g(x)$ approximates $\tilde{\mu}(x)$, i.e. $|g(x) - \tilde{\mu}(x)| \leq \varepsilon$, within the interval $-4A \leq x \leq 4A$, then $g(\pm 4A) \leq 1 - 3\varepsilon$ and $g(0) \geq 1 - \varepsilon$ (Fig.A1). Therefore, there exist such points $x_1 \in [-4A, 0]$ and $x_2 \in [0, 4A]$, that $2\varepsilon \leq g(0) - g(-4A) = g'(x_1)4A$ and $-2\varepsilon \geq g(4A) - g(0) = g'(x_2)4A$, i.e. $g'(x_1) \geq \varepsilon/(2A)$ and $g'(x_2) \leq -\varepsilon/(2A)$. By the same logic, there exists $x_0 \in [x_1, x_2]$, such that $\varepsilon/A \leq |g'(x_2) - g'(x_1)| = |g''(x_0)|(x_2 - x_1) \leq |g''(x_0)|8A$. Therefore,

$$A^2 |g''(x_0)| \geq \varepsilon/8, \tag{A2}$$

and hence the second derivative of any such approximating function $g(x)$ cannot be of the order of $\varepsilon^2/A^2$ everywhere (at least when $\varepsilon < 1/8$).

Let us now show that the Conjecture cannot hold under the conditions of the example from Fig.A1. Let us assume for the moment that it were possible to represent the function $\tilde{\mu}(x)$ from the above example as a sum $\tilde{\mu}(x) = \bar{\mu}(x) + \Delta\mu(x)$, where $\Delta\mu(x)$ were small, i.e.

$\Delta\mu(x) = O(\varepsilon)$, and $\bar{\mu}(x)$ were linearizable, i.e. $\bar{\mu}(x \pm 2h) = \bar{\mu}(x) \pm 2h\,\bar{\mu}'(x) + O(\varepsilon^2)$ for all $x$ and all $h$, such that $|h| \leq A$, and $2h\bar{\mu}'(x) = O(\varepsilon)$. Then we could suitably approximate $\bar{\mu}(x)$ by an infinitely differentiable function, e.g. by $g(x) = \dfrac{1}{\sigma(2\pi)^{1/2}} \int \bar{\mu}(x-y) \exp[-y^2/(2\sigma^2)] dy$. Equation (A1) implies that if $\sigma < \varepsilon A$, then $g(x) - \bar{\mu}(x) = O(\varepsilon)$, and hence $g(x) - \tilde{\mu}(x) = O(\varepsilon)$. It is also easy to see that the function $g(x)$ is still linearizable in the same sense as $\bar{\mu}(x)$. Therefore, $O(\varepsilon^2) = g(x+2h) - g(x) - g'(x)2h = 2g''(x)h^2 + O(h^3)$ when $h \to 0$. However, we have shown above that for any function $g(x)$ which approximates $\tilde{\mu}(x)$ within the strip $[\tilde{\mu}(x) - \varepsilon, \tilde{\mu}(x) + \varepsilon]$ there exists a point $x_0$ where $g''(x_0)$ is not of the order of $\varepsilon^2$, hence the last equation cannot hold for arbitrary small $\varepsilon$. In other words, the function $g(x)$ cannot be linearized in the vicinity of the point $x_0$ at which $g(x)$ is known to have a high curvature. The contradiction that we obtained shows that the Conjecture cannot be true.